\documentclass[pre,aps,preprint,showpacs]{revtex4} 
\usepackage{revsymb,amsmath}
\usepackage{graphicx}
\begin{document}
\author{Steffen Trimper, Knud Zabrocki}
\affiliation{Fachbereich Physik, Martin-Luther-Universit\"at,D-06099 Halle Germany}
\email{trimper@physik.uni-halle.de}
\author{Michael Schulz}
\affiliation{Abteilung Theoretische Physik, Universit\"at Ulm\\D-89069 Ulm Germany}
\email{michael.schulz@physik.uni-ulm.de}
\title{Memory-Controlled Annihilation Reactions}
\date{\today }

\begin{abstract}

We consider a diffusion-limited reaction in case the reacting entities are not available 
simultaneously. Due to the fact that the reaction takes place after a spatiotemporal accumulation 
of reactants, the underlying rate equation has to be modified by additional non-local terms.
Owing to the delay effects a finite amount of reactants remains localized, preventing a further 
reaction and the asymptotic decay is terminated at a finite density. The resulting  
inhomogeneous non-zero stationary concentration is stable against long wave length fluctuations.  
Below a critical wave vector $k_c$ the system becomes inert, whereas a complete decay is realized above $k_c$. 
The phase diagram for the one species-annihilation process $A + A \to 0$ exhibits a behavior comparable 
to a second order phase transition. Obviously the memory effects are equivalent to long range interaction and 
the non-local kinetics is basically independent on space dimensions.
\end{abstract}
\maketitle

\section{Introduction}

In case the typical transport time $\tau_D$ of reacting entities is much 
larger than the reaction time $\tau_R$, the reaction process is affected by spatial 
correlations. Consequently the global concentration $c(t)$ has to be replaced by a local 
density function $c({\bf r},t)$. When the reaction occurs within a cell, centered around the position 
${\bf r}$, the classical mean-field rate evolution equation is completed by a diffusive term,  
which disposes of low dimensional fluctuation corrections. The influence of such fluctuations on the 
long time behavior had been demonstrated in a series of papers \cite{wc}, for a recent review see \cite{0}. 
The newly emerging nonequilibrium behavior is extracted using more sophisticated methods as field theoretic 
treatment \cite{ut} or extensive numerical simulations \cite{h}. In the one species-annihilation process a 
particle $A$ is annihilated upon encounter according to the reaction scheme $A + A \to 0$. The reaction, realized 
with a certain rate $u$, takes place immediately. The time evolution of the local concentration $c({\bf r}, t)$ 
satisfies the rate equation 
\begin{equation}
\frac{\partial c({\bf r},t)}{\partial t} = D\nabla^2 c({\bf r},t) -u c^2({\bf r},t) 
\label{ev}
\end{equation}
Suppose that $l$ is a typical length scale until reactants meet and $\bar{c} \simeq l^{-d}$ a typical 
concentration in that region, the condition $\tau_D \gg \tau_R$ 
leads to 
\begin{equation}
l^{2-d} \gg \frac{D}{u} 
\label{sc}
\end{equation}
Since the diffusion constant $D$ and the reaction rate $u$ become fixed by microscopic processes,  
the spatial variation of the concentration on the length scale $l$ should be relevant for $d < 2$. 
Due to the short reaction time $\tau _R$ the particles annihilate very rapidly. Therefore a subsequent 
reaction can only take place in turn the particles has passed through the distance $l$. This effect slows 
the density decay down. Asymptotically one finds $c(t) \sim t^{-d/2}$ for $d < 2$ and logarithmic 
corrections at the critical dimension $d_c = 2$ \cite {lee}. Recently, the result has been generalized to 
multispecies pair annihilation leading to a modified exponent for the asymptotic decay \cite{tu}. In 
the opposite case $d > 2$, spatial fluctuations are irrelevant and the mean-field equation provides a 
valid description. The diffusion-limited reaction Eq.~(\ref{ev}) is based on the assumption 
that the reactants are available simultaneously. In the present paper we emphasize, that the generic behavior 
of the system may be changed when additional non-linear delay effects are included into the consideration. 
In particular, we discuss a reactions which takes place after a sufficient accumulation time 
and after cumulating reactants within a spatial region. Under that conditions the reaction time scale 
$\tau _R$ is modified and likewise Eq.~(\ref{sc}) is changed. 
Accordingly we demonstrate, the long time behavior is dominated apparently by the delay effects. 
The system is able to reach a stationary state with a finite concentration instead of exhibiting 
an algebraic decay in time as it follows using Eq.~(\ref{ev}).\\ 
Although the crucial factors governing the dynamics of systems, comprising many "units", consists 
of interaction and competition, there is an increasing interest to include memory effects as a further 
unifying feature of complex physical \cite{6a,6b} as well as biological systems \cite{6c}. Our model 
can be grouped into the effort to discuss delay and feedback mechanism. It is well-known that evolution 
equations with memory kernels can be derived following the well established projector formalism due to \cite{mori}, 
see also \cite{naka}. That approach had been applied successfully for the density-density correlation function 
studying the freezing processes in undercooled liquids \cite{l,goe}. Recently a Fokker-Planck equation with a 
non-linear memory term was used to discuss anomalous diffusion in disordered systems \cite{ss}. The results 
could be confirmed by numerical simulations including diffusion on fractals \cite{bst}, 
see also \cite{t,mo}. Moreover, it was argued \cite{loc} that mobile particles 
remain localized due to the feedback-coupling. Notice that a formal solution of 
whole class of non-Markovian Fokker-Planck equations can be expressed through the solution 
of the Markovian equation with the same Fokker-Planck operator \cite{s}. The non-Gaussian 
fluctuations of the asset price can be also traced back to memory effects \cite{st}. 
An additional cumulative feedback coupling within the Lotka-Volterra model, which may stem 
from mutations of the species or a climate changing, leads to a complete different 
behavior \cite{zab} compared to the conventional model. If the Ginzburg-Landau model 
for the time evolution of an order parameter is supplemented by a competing memory term,  
the asymptotic behavior and the phase diagram is completely dominated by such a term 
\cite{zab1}. Whereas the feature of the approach, proposed in those papers, consists of self-organization, 
i. e. the time scale of the memory is determined by the relevant variable itself, for instance the 
concentration, there is a broad class of models with external delay effects \cite{oy,ge,fe}, for a survey 
and applications in biology see \cite{mur}. That case is characterized by a given external memory 
kernel. The differences of both approaches will be discussed also in \cite{zab2}. The spreading of an agent 
in a medium with long-time memory, which can model epidemics, is 
studied in \cite{gr}. Time-delayed feedback control is an efficient method for stabilizing 
unstable periodic orbits of chaotic systems \cite{p}. Time delay may induce various 
patterns including travelling rolls, spirals and other patterns \cite{j}. 
The influence of a global feedback is studied recently in a bistable system \cite{sa}, where the   
purpose of that paper is a discussion of the domain-size control by a feedback. Even in an 
open quantum system non-Markovian dynamics is characterized by time-non-locality in 
the equation of motion for the reduced density operator \cite{mm}.\\ 
In view of the large variety of systems with feedback couplings it seems to be worth to study 
simple models, which still conserve the crucial dynamical features of evolution models as 
non-linearities and moreover, as a new ingredient, delayed feedback-coupling. In the 
present paper we discuss the influence of a non-Markovian memory term on chemical reactions. 
The retardation effects are characterized by memory kernels, which are chosen in such a manner, that 
they compete with the conventional diffusion and the instantaneous non-linear reaction terms. 

\section{Model} 

Although a chemical reaction is characterized by non-linear reaction terms, as it is indicated in Eq.~(\ref{ev}), 
the reaction process may be further modified when the reaction takes place after an accumulation time. In that 
case the time evolution of the concentration could also depend on the history of the sample to which it belongs, i. e. 
the changing rate of the concentration should be influenced by the changing rate in the past. Thus the 
evolution for global concentration $c(t)$ has to be supplemented by memory terms. Such term model, for instance the way 
on which a seed concentration had been accumulated by a delayed transport mechanism, originated 
by the environment of the reactants. With other words, the changing rate of a certain 
quantity at time $t$ is also determined by the accumulation rate at a former time 
$t^{\prime} < t$. In between, i.e. within the interval $\tau = t - t^{\prime}$, the 
reactants are enriched while changing the concentration at $t^{\prime}$. Regardless 
that process and further fluctuations the available amount of concentration 
at time $t$ is governed by instantaneous loss term as well as on 
the changing rate at former times $t^{\prime}$. Consequently the evolution  
Eq.~(\ref{ev}) should be modified according to 
\begin{equation}
\partial _{t}c({\bf r},t) = \nabla^2 c({\bf r},t) - u c^2({\bf r},t) 
-\int\limits_{0}^{t}dt^{\prime }\int\limits_{-\infty }^{\infty }d{\bf r}^{\prime }
K({\bf r}-{\bf r}^{\prime },t-t^{\prime }; c) 
\label{ev1}
\end{equation}
As the main ingredient we assume that the kernel $K$ is determined by the concentration $c$ and its derivative. 
With other words, the time and the spatial scale of the memory is governed by the spatiotemperal scale of the 
concentration. Under that conditions the chemical reaction becomes obviously a many-body problem. The reaction 
species are embedded into an environment of all the other particles of the system which give rise to feedback and 
memory effects.\\
Let us now specify the memory kernel introduced in Eq.~(\ref{ev1}) by the expression
\begin{eqnarray}
\partial _{t}c({\bf r},t) &=& D \nabla^2 c({\bf r},t) - \mu \int\limits_{0}^{t}dt'
\int\limits_{-\infty }^{\infty }d{\bf r'} 
\nabla c({\bf r - r'}, t - t') \cdot \partial_{t'} \nabla c({\bf r'},t') \nonumber\\
&-& u c^2({\bf r}, t) - \lambda  \int\limits_{0}^{t}dt'\int\limits_{-\infty }^{\infty }d{\bf r'} 
c({\bf r - r'}, t - t') \partial_{t'}c({\bf r'},t')
\label{ev2}
\end{eqnarray}
The second term on the right hand side characterizes a delayed diffusion with diffusion parameter $\mu $. 
This term conflicts to the conventional local diffusive one with the parameter $D$. Large concentration fluctuations 
at both times $t$ and $t'$ contribute to the behavior of the system. Different to the constant $D$ the sign 
of the parameter $\mu$ can range between positve and negative values. Likewise the third and the fourth term 
exhibit an extra competition between the local and the accumulated as well as the delayed reaction, where we 
assume $u \geq 0$ and $\lambda$ arbitrary posisitive and negative. Using scaling analysis it is easy to verify 
that the last term with the coupling parameter $\lambda $ is relevant in all dimensions. In contrast the spatial 
dependence of the local $u$-term is only relevant for $d < 2$. The diffusion term and the delayed diffusion term 
are of the same order of magnitude. Follwing the line resulting in Eq.~(\ref{sc}), we can estimate the characteristic 
time scales. Whereas the diffusion time remains unchanged, $\tau_D \simeq l^2/D$, there occurs an effective reaction 
time $\tau _R $ which is composed of the parameters $\mu,\, u$ and $\lambda $ of the three non-linear processes,  
introduced in Eq.~(\ref{ev2}). The effective reaction time is fixed by 
\begin{equation}
\tau _R \simeq \frac{1}{l^d \lambda + l^{d-2} \mu} \left[ u + \sqrt{u^2 + l^d(  l^d \lambda + \mu l^{d-2})}\right]
\label{sc1}
\end{equation}
The condition $\tau_ D \gg \tau _R $ yields 
\begin{equation}
\lambda l^4 + \mu l^2 + l^{2-d} D u \gg D^2. 
\label{sc2}
\end{equation}
In the special case $\lambda = \mu = 0$ this relation leads anew to Eq.~(\ref{sc}). For non-zero memory 
parameters $\mu$ and $\lambda$ Eq.~(\ref{sc2}) is fulfilled in each space dimension. Insofar, the memory 
mimics a kind of very long-range forces. In the forthcoming section, we search for the stationary solutions 
in arbitrary dimensions $d$.

\section{Stationary concentration}
\begin{figure} [!ht]
\centering
\includegraphics[scale=1.2]{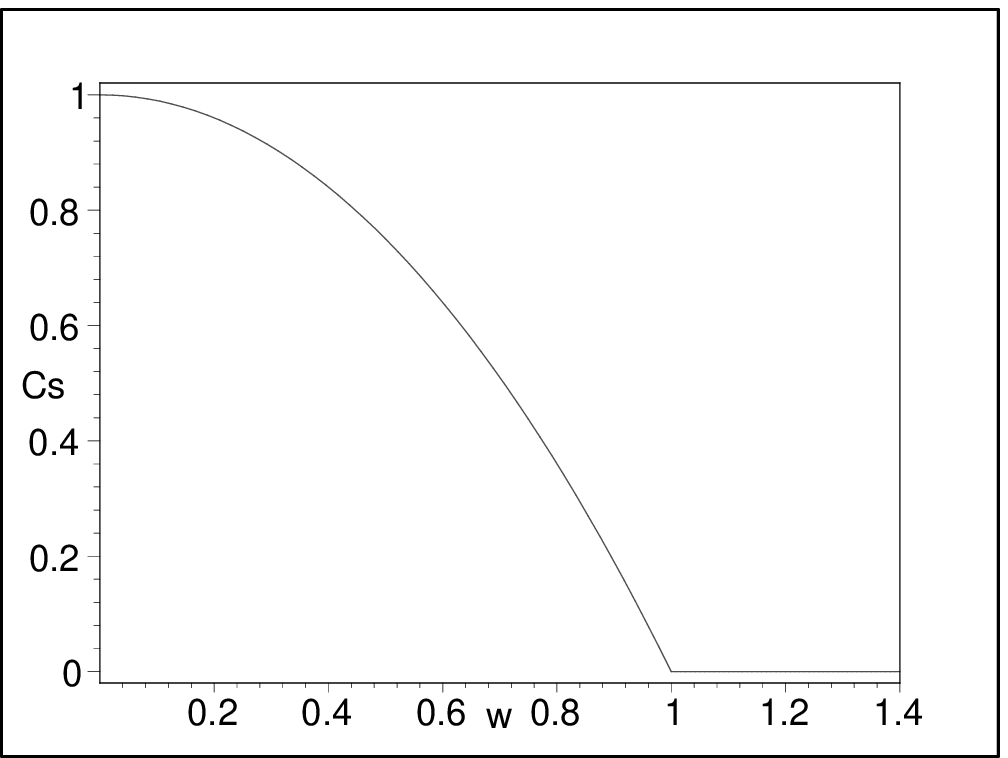}
\caption{Stationary concentration $c_s({\bf k})$ versus $w = k/k_c$ for positve parameters $\mu, \lambda  > 0$ }
\label{fig.1}
\end{figure}

\begin{figure}[!ht]
\centering
\includegraphics[scale=1.2]{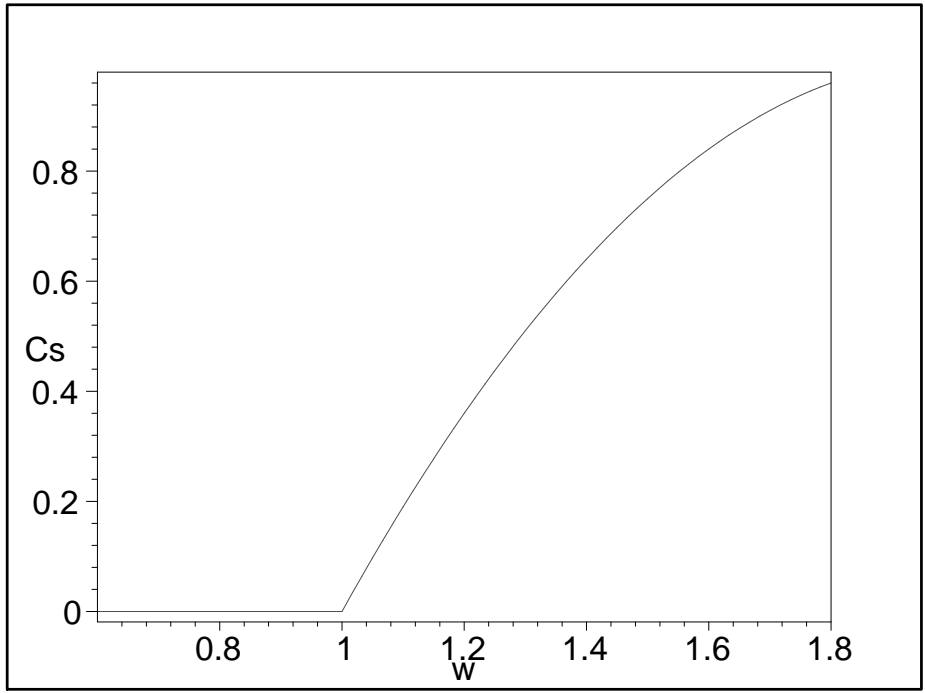}
\caption{Stationary concentration $c_s({\bf k})$ versus $w = k/k_c$ for negative parameters $\mu < 0$ and $\lambda < 0$ }
\label{fig.2}
\end{figure}

In this section we show that the asymptotic behavior of the concentration is changed significantly when 
the retardation effects are taken into account. Firstly let us consider the case of vanishing memory 
parameters $\mu = 0$ and $\lambda = 0$. In mean-field approximation the density decay obeys $c \sim 1/u t$. 
For $d < 2$ the density fluctuations leads to an additional time dependence of the parameter $u \to u(t)$. 
The explicit time dependence of $u$ can be estimated easily by scaling arguments. The parameter $u$ scales 
as $u \sim l^{d - z} $ where the dynamical exponent $z $ is related to the time scale by $l \sim t^{1/z}$. 
Because for conventional diffusion the dynamical exponent is $z = 2$ we find $u(t) \sim t^{d/2 - 1}$ which leads 
immediately to the asymptotic relation $c(t) \sim t^{-d/2}$ in accordance with more refined methods \cite{lee}, 
compare also \cite{0}. All particles are annihilated due to the reaction. This picture is changed, when 
the delay effects are enclosed into the consideration. The asymptotic behavior in time is obtained by applying 
Laplace transformation. Moreover let us perform Fourier 
transformation with respect to the spatial coordinate ${\bf r}$. The concentration 
after both transformations is denoted by $c( {\bf k}, z)$. Making the ansatz 
$c({\bf r}, t)=g({\bf r})+ \varphi ({\bf r},t)$ or 
\begin{equation}
c( {\bf k}, z) = \frac{c_s({\bf k})}{z} + \varphi ({\bf k}, z) 
\label{lt}
\end{equation}
where $\varphi ({\bf k},z) = \mathcal{L}\{\varphi ({\bf k}, t)\} = \int_0^{\infty} \varphi ({\bf k}, t)\} \exp(-zt)dt $ 
should not display a singular behavior for $z \to 0$. Inserting Eq.~(\ref{lt}) into Eq.(\ref{ev2}) the singular 
part yields the stationary solution $c_s({\bf k})$ which obeys the equation  
\begin{eqnarray}
c_s({\bf k}) &=& \frac{c_s({\bf k}) [ (\lambda - \mu k^2) c_0({\bf k}) ] - u b({\bf k})}
{D k^2 + (\lambda - \mu k^2) c_s({\bf k})} \nonumber\\
\mbox{with}\quad b({\bf k}) &=& \int \frac{d^dq}{(2\pi )^d} c_s({\bf q}) c_s({\bf k - q}),
\quad c_0({\bf k})  \equiv c({\bf k}, t  = 0)  
\label{as}
\end{eqnarray}
This relation has always the solution $c_s({\bf k}) = 0$ which corresponds to a complete 
decay, ending up with zero concentration. In addition to this trivial solution Eq.~(\ref{as}) 
exhibits a non-zero stationary solution, although the instantaneous decay process, manifested 
the decay rate $u \neq 0$, is still present. Obviously the memory effects prevent a complete reaction. 
The non-trivial solution of Eq.~(\ref{as}) depends on the initial value $c_0({\bf k})$ 
and on the parameters of the model. A linear stability analysis in terms of $\varphi ({\bf k}, t)$, Eq.~({lt}), leads to
\begin{eqnarray}
\dot{\varphi}({\bf k}, t) &=& - \Lambda ({\bf k}) \varphi ({\bf k}, t) - 2 u \int_q c_s({k - q}) \varphi ({\bf q}, t) 
\nonumber\\\mbox{with}\quad \Lambda ({\bf k}) &=& D k^2 + (\lambda - \mu k^2) c_s({\bf k})  
\label{stab}
\end{eqnarray}
To discuss the generic behavior let us firstly study the case $u = 0$ which is reasonable, since the 
inhomogeneous stationary solution is originated mainly by the memory terms proportional to the coupling 
parameters $\mu $ and $\lambda $, respectively. In that case the nontrivial stationary solution reads in 
according to Eq.~(\ref{as}) 
\begin{equation}
c_s({\bf k};\,u=0) = c_0({\bf k}) - \frac{D k^2}{\lambda - \mu k^2}
\label{8}
\end{equation}
If $\lambda = 0$ it results a homogeneous solution as expected whenever the initial concentration is 
likewise homogeneous. In general the stationary solution is only accessible when $c_s({\bf k}) \ge 0$ and 
reasonable if $c_s({\bf k};\, u = 0) \leq c_0({\bf k})$. Because the asymptotic behavior depends essentially 
on the signs of the feedback parameters $\mu $ and $\lambda$ we consider firstly the case of 
$\mu \geq 0$ and $\lambda \geq 0$. For positive parameters $\mu $ and $\lambda > 0$, the inhomogeneous stationary 
solution is stable and reasonable for $k < k_c$ where the critical wave vector is determined by 
\begin{equation}
k^2_c = \frac{\lambda c_0(k_c)}{D + \mu c_0(k_c)}
\label{9}
\end{equation}  
A simple realization is given for $c({\bf r},\,t=0) = c_0 \delta ({\bf r})$ leading to $c_0({\bf k}) = c_0$.  
In that case the critical wave vector is simply defined by the last equation where $c(k_c, 0)$ is replaced by 
$c_0$. Remark that the critical value $k_c$ fulfills the relation $k_c < \sqrt{\lambda / \mu}$ 
In this manner the spurious singularity in Eq.~(\ref{8}) is avoided.   
Together with the stability criteria $\Lambda ({\bf k}) > 0$ we get the phase diagram depicted in Fig.\ref{fig.1}. 
The curve is similar to the behavior of a second order phase transition. Introducing the variable $w = k/k_c$ we get
\begin{equation}
c_s(w;\,u=0) = (1 - w^2)\frac{c_0 (D + c_0 \mu )}{ D + c_0 \mu (1 - w^2)}
\end{equation}
One can study the spatial dependence of non-trivial behavior of $c_s({\bf r})$ by making the inverse 
Fourier-transformation. For large $ r = \mid {\bf r} \mid \to \infty $ we get
$$
c_s(r) = c({\bf r}, 0) - \frac{\kappa }{r^{d+2}}
$$
where $\kappa$ is a factor depending on the parameters of the model. The physical behavior of the system 
can be discussed in terms of particles and holes denoted as $A$ and 
$B$ particles, respectively. On a large length scale the final state of the system consists of homogeneous 
distributed $A$-particles, which becomes inert and which are separated from the holes. On a short scale 
the $A$-particles disappear further until they are completely annihilated and only holes leave over. 
The systems exhibit a new kind of phase separation mechanism.\\
In the same manner we can discuss the residual cases with varying signs of the feedback parameters $\mu$ and $\lambda$. 
If both parameters are negative, $\mu < 0$ and $\lambda < 0$, we find the inverse situation compared to 
that one discussed above. A non-zero stable and simultaneously reasonable stationary state appears for 
$ k > k_c$, whereas for $ k < k_c$ all the particles undergo the complete reaction, see Fig.\ref{fig.2}. The system 
exhibits a microphase-separation. It decays in small subunits of extension $l < k^{-1}_c$. If $\mu > 0$ and 
$\lambda < 0$ there exists only the trivial stationary solution $c_s({\bf k}) \equiv 0$. In the remaining 
case, $\mu < 0$ and $\lambda > 0$, we find a stable non-zero stationary solution for all wave vectors 
provided the initial concentration fulfills the condition $c_0(k) > D \mid \mu \mid^{-1} $. In the opposite case 
one observes a similar behavior as it is depicted in Fig.\ref{fig.1}.\\
If the reaction parameter $u$ is non-zero, Eq.~(\ref{as}) can not be solved exactly due to the non-linear 
convolution term. However, the main conclusions remain valid, in particular for a small coupling parameter $u$. 
In a linear approximation with respect to $u$ the stationary solution Eq.~(\ref{8}) is modified in according to 
$c_s({\bf k};\,u = 0) \to c_s({\bf k};\,u) \equiv c_s({\bf k};\,u = 0) - u R({\bf k})$ with
$$
R({\bf k}) = \frac{ \int d^dq/(2\pi )^d)\left[ c_s({\bf q};\,u = 0) c_s({\bf k - q};\,u = 0)\right]}
{ (\lambda - \mu k^2) c_s({\bf k};\,u = 0)}
$$   
Expanding the last expression with respect to $k^2$ by using Eq.~(\ref{8}) we find, that the 
critical wave vector $k_c$, compare Eq.~(\ref{9}), is slightly reduced and the stability of the solution is maintained. 
The phase diagram, presented in Fig.\ref{fig.1}, is preserved. 

\section{Conclusions}

In this paper we have extended the conventional modeling of chemical reactions by including non-Markovian 
memory terms within the evolution equation. The additional terms give rise to a competitive behavior 
for both the local reaction and the diffusive transport term. The influence of the feedback-couplings 
is essential, in particular in the long time limit. Due to the memory effects the reactants may be localized and 
therefore a further reaction is prevented. The resulting asymptotic inert state is stable on a large length scale. 
However, there exists a critical wave vector, above it, the reaction can be continued as long as all the reacting 
particles are annihilated completely . The reason for such a new behavior is by means of an explicit coupling of 
the rate of the concentration at the observation time $t$ to that one at a previous time. This time accumulation 
is further accompanied by an additional spatial accumulation, the effect of which is comparable to the effect 
a long-range interaction forces and consequently the results are basically independent on spatial 
dimensions in according to scaling arguments. These many-body effects are shown to change the asymptotic 
behavior drastically. Due to the feedback-coupling of a particle to its environment, a subsequent particle, 
undergoing a diffusive motion, gains information from a modified environment. Hence the particle can be 
confined within a certain region preventing a further reaction. In this manner a self-organized memory leads 
to a non-zero stationary concentration $c_s(r)$, where the analytical form of $c_s$ is controlled by the 
memory strength. This situation is realized on a large length scale. Below a critical wave vector $k_c$ the 
inhomogeneous stationary solution is stable. The phase diagram is similar to an equilibrium 
phase transition of second kind. The model could be relevant for very complex chemical reactions where not all 
entities are present simultaneously.

\begin{acknowledgments} 
This work was supported by the DFG (SFB 418).
\end{acknowledgments}

\newpage

\end{document}